\documentclass[pre,twocolumn,amsmath,amssymb,showpacs]{revtex4}
\usepackage{wasysym}
\usepackage{epsf}
\usepackage{graphicx}
\usepackage{bm}

\usepackage{graphicx}

\usepackage{amssymb}
\usepackage{bm}
\usepackage{amssymb}
\def\ZZ{{\mathbb Z}}

\begin{document}



\title{Why dynamos are prone to reversals}



\author{F. Stefani, G. Gerbeth, U. G\"unther, M. Xu}

\address{Forschungszentrum Rossendorf, P.O. Box 510119,
D-01314 Dresden, Germany}

\begin{abstract}
In a recent paper \cite{PRL} it was shown that
a simple  mean-field dynamo model with a spherically symmetric
helical turbulence parameter $\alpha$
can exhibit a number of features which are
typical for Earth's magnetic field reversals.
In particular, the model
produces asymmetric
reversals (with a slow decay of the dipole of
one polarity and a fast recreation of the
dipole with opposite
polarity), a positive correlation of field strength and
interval length, and a bimodal field
distribution.
All these features are attributable
to the magnetic field
dynamics in the vicinity of an
{\it exceptional point} of the spectrum
of the non-selfadjoint dynamo operator where two real
eigenvalues
coalesce and continue as a complex conjugated pair of
eigenvalues. Usually, this exceptional point is associated with
a nearby local maximum of the
growth rate
dependence on the magnetic Reynolds number.
The negative slope of this curve between the local
maximum and the exceptional point makes the
system unstable and drives it
to the exceptional point and beyond into the
oscillatory branch where the sign change happens.
A weakness of this reversal model
is the apparent necessity to fine-tune the
magnetic Reynolds number and/or
the radial profile of $\alpha$
in order to adjust the operator spectrum in an appropriate way.
In the present paper, it is shown that
this fine-tuning is not necessary in the case of higher
supercriticality of the dynamo.
Numerical examples and physical arguments
are compiled to show that, with increasing magnetic
Reynolds number,
there is strong tendency for the exceptional point and the
associated local maximum to
move close  to the zero growth rate line where the
indicated reversal scenario can be actualized.
Although exemplified again
by the spherically symmetric $\alpha^2$ dynamo model,
the main idea of this ''self-tuning'' mechanism
of saturated dynamos into a reversal-prone state
seems well transferable to other dynamos.
As a consequence, reversing dynamos might be much more
typical and may occur much more frequently in nature
than what could be expected from a purely
kinematic perspective.

\end{abstract}


\maketitle

\section{Introduction}

The magnetic field of the Earth is
known to undergo irregular
reversals of its dipole part. The reversal rate
is variable in the
course of time: it
was nearly zero in the Kiaman
and the Cretaceous
superchrons and is
approximately 5 per Myr in the present \cite{MERR}.

Much effort has been devoted
to identify typical characteristics of the
reversal process.
In particular, it was claimed that reversals
may have an asymmetric, saw-toothed
shape, with
the field of one polarity decaying slowly and recreating
rapidly with opposite polarity, possibly to quite
high intensities
\cite{VALET,MEYN,BOGU}.

Another hypothesis concerns a
correlation between the polarity
interval length and the
magnetic field intensity \cite{COX,TARDUNO}.
Associated with this,
a causal connection of the Cretaceous superplume and
superchron period has been discussed \cite{FULLER,LARSON}.
It was Vogt who first suggested a correlation between
volcanism and reversal rate \cite{VOGT}. The general
idea behind this is that superplumes give rise to an
increased heat transport
from the core mantle boundary to the Earth surface
with the result of an increased dynamo
strength due to a higher
temperature gradient driving the outer core flow
\cite{LARSOLS}.
While this idea was soon generally accepted (with some
counter-arguments regarding the involved
time-scales \cite{LOPER}), quite
contrary implications for
the reversal frequency were drawn from it.
The first ''school'', advocating
a negative correlation of
interval length and
energy supply to the dynamo,
goes back to Loper and McCartney \cite{LOMC}.
The second school, suspecting long intervals
for a
strong dynamo, was motivated by
various mean-field dynamo models,
for which a transition from
anharmonic oscillations to superchrons
for increasing magnetic Reynolds number was observed
\cite{OLSLEE}.

A third, and still controversially discussed
observation concerns the {\it bimodal distribution}
of the Earth's virtual dipole moment (VDM) with two peaks
at about 4 $\times$ 10$^{22}$ Am$^2$
and at about twice that value \cite{PERRIN,SHCH,HELLER}.

For decades, it has been a challenge for dynamo
theoreticians to explain reversals and their
characteristics. It was considered a breakthrough when
Glatzmaier and Roberts observed a
reversal process in their fully coupled three-dimensional
simulation of the geodynamo \cite{GLRO}
(cf. also \cite{GLRO-ANDCO} for a recent
overview).
The strange thing with these simulations is that they
reproduce many features of Earth magnetic fields,
including reversals, quite well despite the
fact that they are
working in parameter regions far beyond those of the real
Earth. This deficiency applies, in particular,
to the Ekman
and the magnetic Prandtl number. A way out of this
dilemma may lie with
a reliable sub-grid scale modeling
\cite{MININNI}.
In this respect one should also notice
recent efforts to link direct numerical
simulations and
mean-field dynamo models \cite{GIES,SCHRINNER}.

This brings us back from expensive simulations
to the complementary tradition of understanding
reversals in terms of reduced dynamo models.
A very simple approach in this direction is the
celebrated Rikitake dynamo  of two coupled
disk dynamos \cite{RIKI,FRANCK}.

Another model was studied by Hoyng and collaborators
\cite{HOYNG,SCHMITT,HOYNG2}. A mean-field dynamo model
is reduced to an equation system for
the amplitudes of the non-periodic axisymmetric dipole mode
and for one periodic overtone under
the influence of stochastic forcing.
This simple model, which produces
sudden reversals and a
Poissonian distribution of the interval time,
has also
been employed to simulate the phenomenon of
stochastic resonance
\cite{LORITO}.
Stochastic resonance
was made responsible in a former paper
\cite{CONSOLINI}
for an apparent 100 kyr periodicity in the interval length
distribution  \cite{YAMAODA}
(note however, that this periodicity is not settled yet
\cite{WINKL}).
An essential ingredient of Hoyng's  model to explain
the correct reversal duration and the interval
length consistently
is the use
of a large turbulent resistivity which is hardly
justified.  At least nothing of this has been seen
in the recent liquid sodium dynamo
experiments \cite{RMP}.

A further approach to understand reversals
relies on
the transition between non-oscillatory
and oscillatory eigenmodes of the dynamo operator
\cite{PARKER,YOSHI,SAJO}.
Those transition points, which have been found in
many dynamo models {\cite{DEIN,DUDL},
are well known in operator theory as spectral branch points ---
''exceptional points'' of branching type of non-selfadjoint operators
\cite{KATO}. Such branch points are characterized not only
by coalescing
eigenvalues but also by a coalescence of two or more (geometric)
eigenvectors and the formation of a non-diagonal Jordan block structure
with associated vectors (algebraic eigenvectors) \cite{CZECH1,SEYR,CZECH2}.
This
is in contrast to ''diabolical points'' \cite{BERRY} which
are exceptional points
of an accidential crossing of two or more spectral branches with an
unchanged diagonal block structure of the operator and without
coalescing eigenvectors  \cite{KATO,SEYR}.

In a recent paper \cite{PRL}, we have analyzed
the magnetic field dynamics
in the vicinity of an exceptional point in more detail.
Although the used model, a mean-field dynamo of the
$\alpha^2$ type with a supposed spherically
symmetric helical turbulence paramter $\alpha$,
is certainly far beyond the
reality of the
Earth dynamo (owing, in particular, to the missing
North-South-asymmetry of $\alpha$)
it exhibited all mentioned reversal
features: asymmetry,
a positive
correlation of field strength and interval
length, and
bimodal field distribution.

All those features together were attributed
to the very
peculiar magnetic field
dynamics in the vicinity of an exceptional point.
Usually, this exceptional point is associated with
a local maximum of the
growth rate curve at a slightly lower
magnetic Reynolds
number.

If this local maximum lies above zero,
then there is a stable fixed point to the left of it.
However, any prevailing noise can trigger the system
to switch to
the unstable fixed point at the right of the
local maximum.
From there the system is driven, in an self-accelerating way,
to the exceptional point and beyond
into the oscillatory
branch where it undergoes the very polarity change and comes
back to one of the fixed points.

If the local maximum is below zero,
the system undergoes an anharmonic oscillation (a limit cycle)
with a pronounced
asymmetry of the ''reversal''. However, noise can
lift the local maximum above zero
making the system stay in the fixed point for a while before
resuming the anharmonic oscillation.

This reversal scenario seems
not unrealistic since
it exhibits at least three reversal characteristics
with only demanding
the existence of an exceptional point and a
nearby local maximum
of the growth rate.
The bad news is that
it seems to require an
artificial fine-tuning of the intensity and/or
the spatial distribution of
the  dynamo source, in order to position
the exceptional point and its accompanying local maximum
close to the zero line.
By checking a variety of $\alpha$ profiles in the {\it
kinematic regime}, we have indeed observed that the
spectral structure as it
is necessary for reversals to happen occurs only seldom.
Hence, the criticism that our particular
choice of $\alpha(r)$
has not enough  geophysical
background to explain reversal \cite{GIES},
seems well justified.

Thus motivated, it is the
main goal of the present paper
to find better arguments why a dynamo operator
should have a reversal-prone spectrum.
A first hint on the solution of this puzzle
can be found in  the papers by Brandenburg
et al. \cite{BRANKRAU}
and by Meinel and Brandenburg \cite{MEINBRAN}.
For a mean-field disc dynamo (which is rather
a model for galactic than for planetary dynamos),
it was shown \cite{MEINBRAN} that a dynamo in {\it the highly
supercritical regime} can exhibit a pronounced reversal
behaviour, although this would not be expected from
considering only the kinematic profiles of $\alpha$.

In the present paper we will verify if this behaviour
in the highly supercritical regime can
be understood in terms of the ''exceptional point model''.
Our main outcome will be
that there is {\it a strong tendency of
saturated dynamos to evolve into a reversal-prone
state} where the condition, that the
exceptional point and the local maximum are situated
close  to the
zero line, is indeed fulfilled.
Applied to the Earth, a possible conclusion could be
that reversals are rather due to a {\it self-tuning}
of the saturated dynamo operator
than to an accidental fine-tuning of the
kinematic dynamo operator.

\section{The model}

Instead of simulating the real Earth dynamo
with a fully coupled three-dimensional solver, we
employ a very simple mean-field dynamo model
in order to work out clearly the basic mechanism of
reversals.

The considered  $\alpha^2$ dynamo with a
spherically  symmetric helical turbulence parameter
$\alpha$ leads to a system of two coupled partial
differential equations with only one spatial variable
(the radius).
This model is simple enough to
allow for long-time simulations providing reasonable reversal
statistics, but at the same time it is still
complex enough to catch the essence of
{\it hydro}magnetic dynamos.
In contrast to their technical counterparts,
the saturation of hydromagnetic dynamos relies strongly
on the {\it deformability of the dynamo source} which is,
in our case, the variable radial dependence of $\alpha$.
Note that in the quite different context
of the Riga dynamo experiment
a similar one-dimensional back-reaction model was
used which reflects also the
hydromagnetic character of this
dynamo experiment in contrast to experiments with
a more constraint geometric flexibility
\cite{PLASMA,MOMOMO}.

The magnetic field evolution of a kinematic
$\alpha^2$ dynamo
is governed by the induction equation
\begin{eqnarray}
\frac{\partial {\bm{B}}}{\partial \tau}={\bm \nabla}
\times (\alpha {\bm{B}}) +
\frac{1}{\mu_0 \sigma} \Delta {\bm{B}} \; ,
\end{eqnarray}
with $\alpha$ denoting the
helical turbulence parameter which may depend on the
position $\bf r$ and
the time $\tau$ \cite{KRRA}. The dynamo acts,
within a sphere of radius
$R$, in
a fluid with
electrical conductivity $\sigma$. The magnetic field has to be
divergence-free, ${\bm{\nabla}} \cdot {\bm{B}}=0$.
In what follows, the length will be measured in units of $R$,
the time in units of $\mu_0 \sigma R^2$,
and
the parameter $\alpha$ in units of
$(\mu_0 \sigma R)^{-1}$.
Note that for the Earth  we get a typical
time scale $\mu_0 \sigma R^2 \sim 200$ kyr,
which results in a free decay time of 20 kyr for
the dipole field.

It is convenient to decompose ${\bm{B}}$ into a
poloidal and a toroidal field component according to
${\bm{B}}=-\nabla \times ({\bm{r}} \times
\nabla S)-{\bm{r}} \times
\nabla T $.
The defining scalars $S$ and $T$ are
then expanded
in spherical harmonics of degree $l$ and order $m$
with the expansion coefficients
$s_{l,m}(r,\tau)$ and $t_{l,m}(r,\tau)$.

For the present case with $\alpha{(\bm r})=\alpha(r)$,
the induction equation
decouples for each $l$ and $m$ into the
following pair
of equations:
\begin{eqnarray}
\frac{\partial s_l}{\partial \tau}&=&
\frac{1}{r}\frac{d^2}{d r^2}(r s_l)-\frac{l(l+1)}{r^2} s_l
+\alpha(r,\tau) t_l \; ,\\
\frac{\partial t_l}{\partial \tau}&=&
\frac{1}{r}\frac{d}{dr}\left[ \frac{d}{dr}(r t_l)-\alpha(r,\tau)
\frac{d}{dr}(r s_l) \right]\nonumber\\&&-\frac{l(l+1)}{r^2}
[t_l-\alpha(r,\tau)
s_l] \; .
\end{eqnarray}
The boundary conditions are
$\partial s_l/\partial r |_{r=1}+{(l+1)} s_l(1)=t_l(1)=0$.
In the following we focus our attention on the dipole
mode with $l=1$, and will henceforth skip the corresponding
subscript of $s$ and $t$; $s:=s_1$ and $t:=t_1$.
The absence of the order $m$ in Eqs. (2,3)
follows from the spherical symmetry of $\alpha$. It implies,
in particular,
a complete degeneration of axial and equatorial
dipole modes. It is clear that for any more realistic model
(e.g. with inclusion of the North-South asymmetry of $\alpha$)
this degeneration would be lifted.

Let us assume that the
profile of $\alpha$ in the kinematic regime, $\alpha_{kin}(r)$,
represents a
supercritical
dynamo.
After self-excitation has occurred,
magnetic field saturation is ensured by quenching the
parameter $\alpha$.
We do this with the angularly averaged
magnetic field
energy which
can be expressed in terms of $s(r,t)$ and $t(r,t)$
\cite{NAKAKONO}.
This averaging over the angles, which represents
a severe simplification, has been introduced in
order to remain within the framework of the
spherically symmetric $\alpha^2$ model.
In reality, of course,
any quenching would introduce terms breaking
the spherically symmetry  of $\alpha$.

In addition to the quenching effect, we
assume the $\alpha$-profile to
be affected by
"blobs" of noise which are
considered constant within a correlation
time $\tau_{corr}$.
Physically, this noise is not unnatural: it could
be understood
as a consequence of changing  boundary
conditions for the
core flow, but also as a shorthand for the omitted
influence of
higher multipole modes on the dominant dipole mode.

Putting all together, $\alpha(r)$ takes on
the time dependent form
\begin{eqnarray}
\alpha(r,\tau)&=&C \; \frac{\alpha_{kin}(r)}{1+
E_{mag}(r,\tau)/E^0_{mag}}+\Xi(r,\tau) \; ,
\end{eqnarray}
where
$E_{mag}$ is the magnetic energy, averaged over the
angles,
\begin{eqnarray}
E_{mag}(r,\tau)=\frac{2 s^2(r,\tau)}{r^2}+
\frac{1}{r^2}\left( \frac{\partial (r s(r,\tau))}
{\partial r} \right)^2
+t^2(r,\tau) \; .
\end{eqnarray}
In the numerical scheme, the noise term $\Xi(r,\tau)$
will be treated in form of a Taylor expansion,
\begin{eqnarray}
\Xi(r,t)=\xi_1(\tau) +\xi_2(\tau)
\; r^2 +\xi_3(\tau) \; r^3+\xi_4(\tau) \; r^4 \; ,
\end{eqnarray}
with  the noise correlation given by
$\langle \xi_i(\tau) \xi_j(\tau+\tau_1)
\rangle = D^2 (1-|\tau_1|/\tau_{corr})
\Theta(1-|\tau_1|/\tau_{corr}) \delta_{ij}$.

In summary, our model is governed by
four parameters, the
magnetic Reynolds number $C$, the noise
amplitude $D$, a mean magnetic energy $E^0_{mag}$
in the saturated regime, and the noise
correlation time $\tau_{corr}$.

The equation system (2)-(4) is time-stepped
using an Adams-Bashforth method.
For the following examples,
the correlation time $\tau_{corr}$ has been set to 0.02,
and $E^0_{mag}$ has been
chosen to be 100. The details of these choices are not
very relevant. Roughly speaking, a shorter correlation
time $\tau_{corr}$ would require a stronger
noise amplitude $D$ in order to yield the same effect.

\section{Anharmonic oscillations in the
vicinity of exceptional points}

In \cite{PRL} we had considered the following
particular
form of the
kinematic $\alpha$ profile:
\begin{eqnarray}
\alpha_{kin}(r)=C (-21.5+426.4 \; r^2-806.7 \; r^3+392.3 \; r^4) .
\end{eqnarray}
Actually, this strange-looking Taylor expansion was
the result of an  Evolution Strategy search for
oscillatory
spherically symmetric $\alpha^2$ dynamos \cite{OSZI}.
For this $\alpha(r)$ profile, the growth rate dependence on $C$
is shown as curve K$_1$ in Fig. 1b.
''K'' stands for kinematic, and the subscript ''1''
indicates the eigenfunction with the radial wavenumber 1.
Correspondingly, K$_2$ denotes the eigenfunction with
the radial wavenumber 2. K$_1$ and K$_2$ coalesce at the
(leftmost) exceptional point E and continue as a pair
of complex conjugate eigenvalues. At the upper
exceptional point E', the curves
K$_1$ and K$_2$ split off again into a pair
of real eigenvalues.

The critical value for this $\alpha(r)$ profile, $C=1$,
lies on the oscillatory
branch, enclosed between the two exceptional points
E and E'. For slightly supercritical values of $C$,
the time evolution is a nearly harmonic
oscillation that  becomes more and more anharmonic and
asymmetric with increasing $C$.

At $C=1.2785$ this oscillation acquires nearly a
rectangular form (Fig. 1a).
For this extremely anharmonic oscillation
we analyze, at the eight different instants 1...8
shown in Fig. 1a,
the corresponding instantaneous profiles
$\alpha(r,\tau)$ (Fig. 2a).
Governed by Eqs. (4) and (5), these instantaneous
profiles
depend on the instantaneous magnetic field variables
$s(r,\tau)$ and $t(r,\tau)$, which are shown in
Figs. 2b and  2c, respectively.
For each of these instantaneous $\alpha$ profiles,
the full squares on the vertical dashed line
in Fig. 1b show
the resulting instantaneous growth rate.
In order to identify the position of these points
with respect to the exceptional point, we
put them in the context of a whole growth rate
curve, by artificially replacing the value $C=1.2785$ of the
actual quenched
$\alpha$ profile by values between 0.7 and 1.4.
This way we obtain the thin lines 1...8
in Fig. 1b which makes it possible
to interprete a reversal
in terms of the consecutive deformation of the original
kinematic growth rate curve.

Let us begin with the instant 1. The magnetic field
(Fig. 2b, 2c)
is high, therefore the quenching of $\alpha$ is also
quite strong (Fig. 2a).
The resulting instantaneous growth rate (Fig. 1b)
sits nearby the local
maximum which is slightly below zero.
Hence, the field starts to decay slowly.

At the instant 2, the field has already weakened,
the quenching of $\alpha$ is less pronounced as  before,
and
the point
of the instantaneous growth rate has
moved in between the
local maximum and the
exceptional point.
At the instants 3 and 4, characterized by even
weaker fields and less quenching of $\alpha$,
the growth rate point
has reached the oscillatory branch.
It follows a short intermezzo in the
higher non-oscillatory branch (instant 5)
where the
$\alpha$ profile has nearly taken on the
unquenched shape.
After this, the system returns via the
oscillatory branch (instant 6) to the reversed state at instants
7 and 8.
It should be noted that this reversal scenario
does not need any dramatic change or even a
total sign reversal of $\alpha$,
as it was
proposed e.g. in \cite{OLSON}.

Readers familiar with the van der
Pol oscillator \cite{VANDERPOL} may
notice that the signal form of Fig. 1a resembles strongly
a ''relaxation oscillation''
(we thank
Clement Narteau for drawing our attention to this point).
Indeed, a closer
inspection of the dynamics of the
van der Pol oscillator
shows a similar behaviour of the leading instantaneous
eigenvalues.

\begin{figure}
\includegraphics[width=7cm]{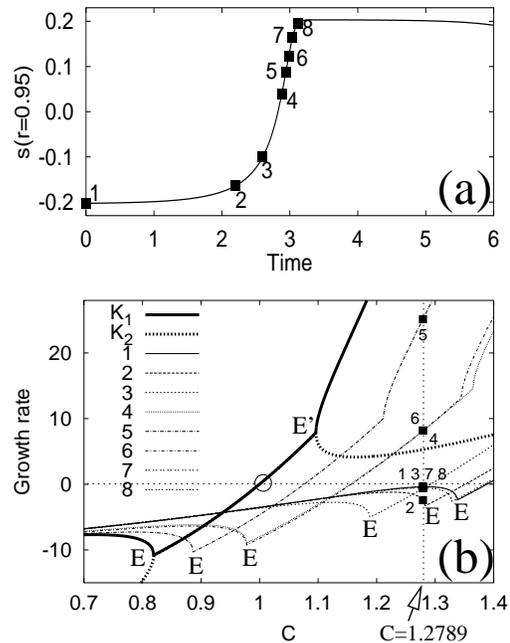}
\caption{(a) Part of an anharmonic oscillation
showing the asymmetry of the reversals for the
case $C=1.2785$. The points 1...8 indicate the instants
which are analyzed in the text.
(b) Growth rates for the kinematic profile
corresponding to Eq. (7) and for the quenched
$\alpha$ profiles at the eight
instants 1...8.
The circle marks the critical point $C=1$ for the kinematic
$\alpha$ profile. ''E'' indicates, for each of the considered
$\alpha$ profiles, the lower exceptional point where the two first
eigenmodes with radial wave numbers 1 and 2
coalesce.
''E' '' marks the second exceptional point,
beyond which the two eigenvalues split and
continue as real ones again.}
\end{figure}
\begin{figure}
\includegraphics[width=7cm]{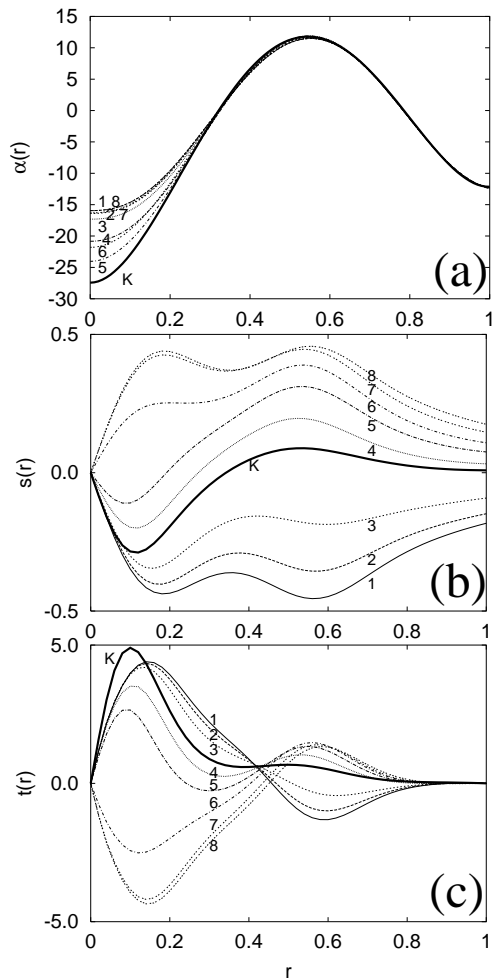}
\caption{Instantaneous $\alpha$ profiles and
defining scalars of the magnetic field at the eight instants indicated
in Fig 1a. (a) $\alpha(r)$. (b) $s(r)$. (c) $t(r)$.}
\end{figure}

At $C=1.2789$, the local maximum
of the growth rate
crosses the zero line, and instead of the considered
{\it limit cycle} in form of an anharmonic oscillation we
get two
fixed points (the crossing points of the growth rate curve
with zero). The fixed point  at the  left of the local
maximum is a stable one. So it will need some noise to
jump over the local maximum. In this regime, the
interval length
is not anymore
governed by the intrinsic frequency of the anharmonic
oscillation,
but by the intensity of the noise which triggers
transition from the stable fixed point to the unstable
fixed point.

\section{Saturation into reversal-prone states}

After the short review of \cite{PRL}
given in the last section, we are
left with the impression that the indicated
reversal scenario
depends heavily on an artificial fine-tuning of the shape and
the intensity of the $\alpha(r)$ profile.

In this section we will try to find better arguments
for dynamos to be in a reversal-prone state.
For this purpose, we change our focus from slightly
supercritical dynamos to highly supercritical dynamos.

One comment is due in
advance. The very particular
$\alpha$ profile in Eq. (7)
was the outcome of an Evolutionary Strategy search for
an
{\it oscillatory and dominant} dipole mode.
In \cite{OSZI} it was shown that this double
demand constrains the
variety of possible $\alpha(r)$ profiles to a rather
thin corridor.
It is much easier to find profiles
$\alpha(r)$  with an
oscillatory $l=1$ mode
for which some higher multipole modes
with $l=2,3,..$ are dominant.
In order not to overcomplicate the problem,
in the following discussion we lift the demand
for a dominant dipole mode by simply
omitting any consideration of higher dipole modes.

Let us start with the classical
profile  $\alpha_{kin}(r)=C$ which is
known to possess only real eigenvalues for all $C$
\cite{KRRA}.
For the sake of concreteness, we consider the
critical value $C=$ 4.49 and
the supercritical values $C=$ 10, 20, and 50.
We solve the induction equation system (2) and (3)
coupled to the
quenching equations (4) and (5).
The quenched $\alpha$ profiles,
$\alpha(r)=C/(1+E_{mag}(r)/E^0_{mag})$,
which are shown in Fig. 3b,
are then
multiplied by a
scaling parameter $C^*$,
and the resulting instantaneous
growth rate
curves in dependence on
$C^*$ are shown in Fig. 3a. As in the former section,
this procedure is intended to
identify the position of the actual growth rate point
with respect to the exceptional point.
\begin{figure}
\includegraphics[width=7cm]{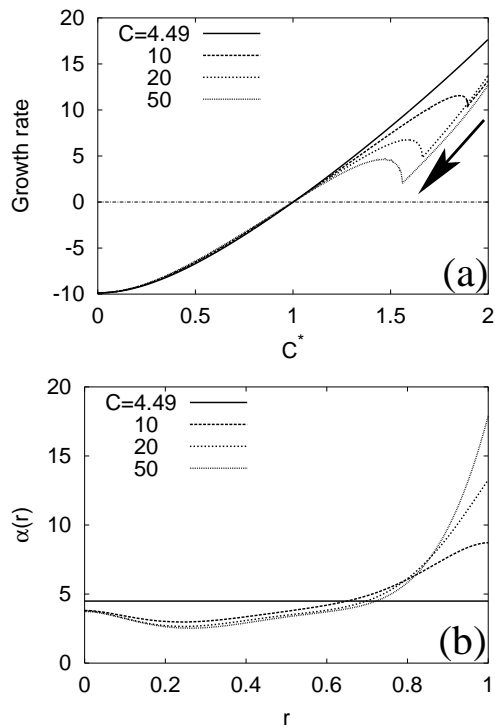}
\caption{(a) Growth rates for the profiles
$\alpha(r)=C^* \cdot C/(1+E_{mag}(r)/E^0_{mag})$ with $C=$10, 20, 50.
The asymptotic quenched state is in all cases non-oscillatory. (b)
Kinematic and quenched profiles $\alpha(r)$ for the four considered
values of
$C$.}
\end{figure}

Fig. 3b shows that the quenching of $\alpha$ is
not homogenous along
the radius.
The saturation mechanism modifies the
original constant $\alpha$ in such a way that
there is a stronger suppression for  smaller
radii (evidently, because
the magnetic field is strongest in the central part
of the dynamo).
This modification of the {\it shape} of $\alpha$
has a remarkable consequence for the spectrum.
In Fig. 3a we see
that the growth rate curves acquire an exceptional
point which is moving
toward the
zero growth rate line with increasing $C$.
In this particular example, the exceptional point
does not drop below
zero. Nevertheless, we will find later  (Fig. 9a) that
even in this case the noise can trigger
reversals.
\begin{figure}
\includegraphics[width=7cm]{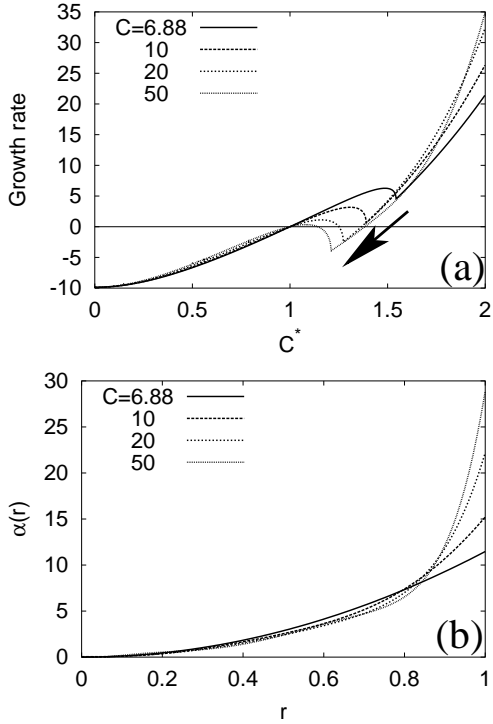}
\caption{Same as Fig. 3, but for
$\alpha(r)=5/3 \; C^* \cdot C r^2/(1+E_{mag}(r)/E^0_{mag})$,
for which the critical value is
$C=6.88$. }
\end{figure}

The next example, $\alpha_{kin}=5/3 \; C\; r^2$, is illustrated in Fig. 4
(the factor 5/3 results from normalizing the radial
average of $\alpha(r)$ to the value for constant $ \alpha$).
The general tendency is the same as for the previous
example $\alpha_{kin}=C$,
but now the
exceptional point drops clearly below the
zero line, leaving the maximum of this curve
only slightly
above zero.
For $C=50$, say, the dynamo will ''sit'' on the stable fixed point
at the left of the local maximum.
At this point the field is rather strong. However, any noise can
trigger a move
to the unstable fixed point at the right of the local
maximum from where the reversal process can start.
\begin{figure}
\includegraphics[width=7cm]{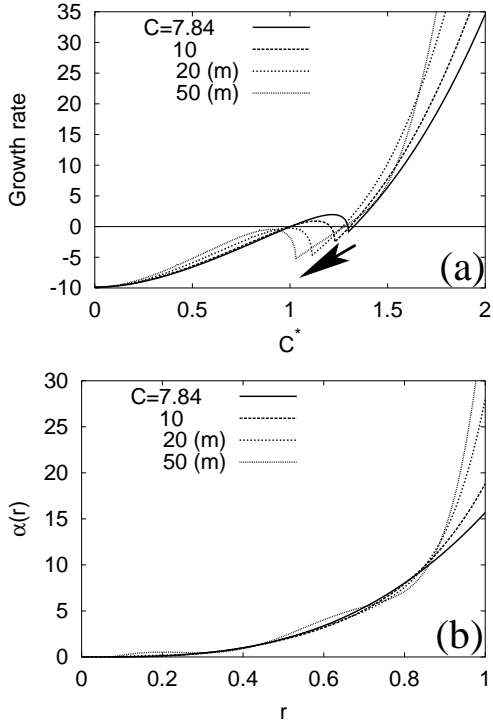}
\caption{Same as Fig. 3, but for
$\alpha(r)=6/3 \; C^* C r^3/(1+E_{mag}(r)/E^0_{mag})$ for which
the critical value is
$C=7.84$. The labels ''20 (m)'' and ''50 (m)'' refer to the
maximally quenched $\alpha$ profiles during
the anharmonic oscillation.}
\end{figure}

The next example is $\alpha_{kin}=6/3 \; C \; r^3$. Here
the local maximum drops below the zero line for $C=$ 20 and 50 (Fig. 5).
That means
there is no stable fixed point anymore, the system runs into a limit cycle
in form of the previously analyzed
anharmonic oscillation. Therefore, $\alpha$ is
changing its shape during this process. The curves ''20 (m)'' and
''50 (m)'' refer to those $\alpha$ profiles which are maximally
quenched during the oscillation.

The last example, $\alpha_{kin}(r)=5.75/3 \;  C \; (1-6r^2+5r^3)$,
is quite similar to the example from \cite{PRL},
but it will provide a surprise (Fig. 6). Starting from the kinematic $\alpha$,
for which the
exceptional point is well below the zero line, it
rises rapidly above zero to a maximum value, but for even higher $C$
it moves back in
direction of the
zero line. Evidently, there are tow different mechanisms at work here.
\begin{figure}
\includegraphics[width=7cm]{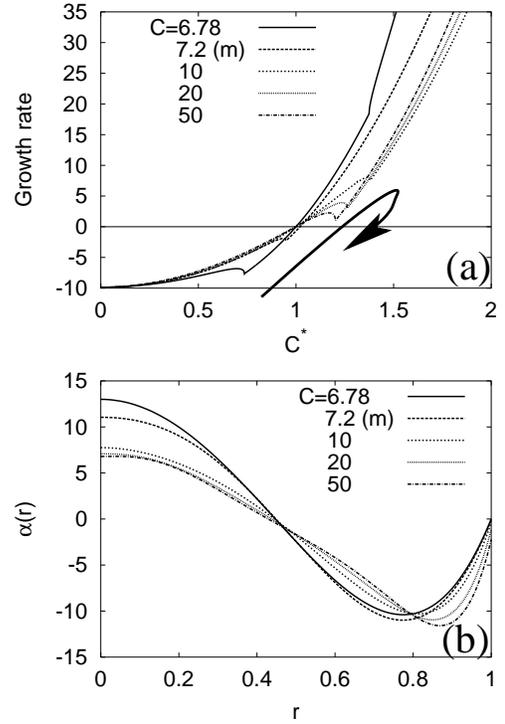}
\caption{Same as Fig. 5, but for
$\alpha(r)=5.75/3 \; C^* \cdot C (1-6r^2+5r^3)/(1+E_{mag}(r)/E^0_{mag})$
for which the critical
value is
$C=6.78$, leading to an oscillatory mode. The label ''7.2 (m)'' indicates
again the maximally quenched $\alpha$. Note the move of the exceptional point
well above the zero line and back to it.}
\end{figure}

All four considered examples exhibit a tendency
for supercritical dynamos to saturate in a state
for which
the exceptional point and its associated local maximum
lie close to the zero growth rate line.
Interestingly enough, this happens independently on whether
the exceptional point in the kinematic case
was above the
zero line (including the limiting case that there was
no exceptional case at all)
or below it.

What is the physical rationale behind this phenomenon?
Back-reaction for hydromagnetic dynamos is quite generally
an actualization of Lenz's rule stating that the excited
magnetic field
acts against the source of its own generation.
For the considered model
this means that
the $\alpha$ profile is deformed in such a way that the growth
rate is
decreasing.
In the first three examples this
makes the growth rate curve of the leading eigenmode
come closer to the
growth rate
curve for the next but leading eigenmode (Fig. 7a).
As a results, the back-reaction
has a
tendency to move the exceptional point
downward in direction of the
zero line (Fig. 7a).

While this argument
is rather intuitive for the first three examples, it does not
apply to the fourth example in which an
exceptional point is already existent {\it below} the zero line.
Again, back-reaction is expected to lower the growth
rate, but now the most efficient way to do this is
quite different.
Figure 7b may help to illustrate the essential point.
The eigenvalues with radial wave numbers
$n=1$ and $n=2$ have already coalesced and formed an oscillatory
eigenmode
below the zero line, but soon after they split again into
a pair of real eigenvalues. The very steep increase of the leading
eigenvalue in this upper real branch is quite typical and is
the heritage of the {\it level repulsion} which would
have occurred for a dynamo operator with a
slightly modified $\alpha$ profile which is closer to the
constant one. Therefore, the most efficient way of
back-reaction in this case is to {\it decrease the steepness} of
this leading
eigenvalue branch. The decrease
of the steepness goes hand in hand with a
lengthening of the oscillatory branch.
This saturation mechanism is so efficient that
the local maximum
and the exceptional point are even risen up towards the
zero line and beyond.
Only later, when the exceptional point is already well
above the zero line, any further increase of $C$ will
promote the previous saturation mechanism,
driving the exceptional point again back to the zero line
(cf. Fig. 6).
\begin{figure}
\includegraphics[width=7cm]{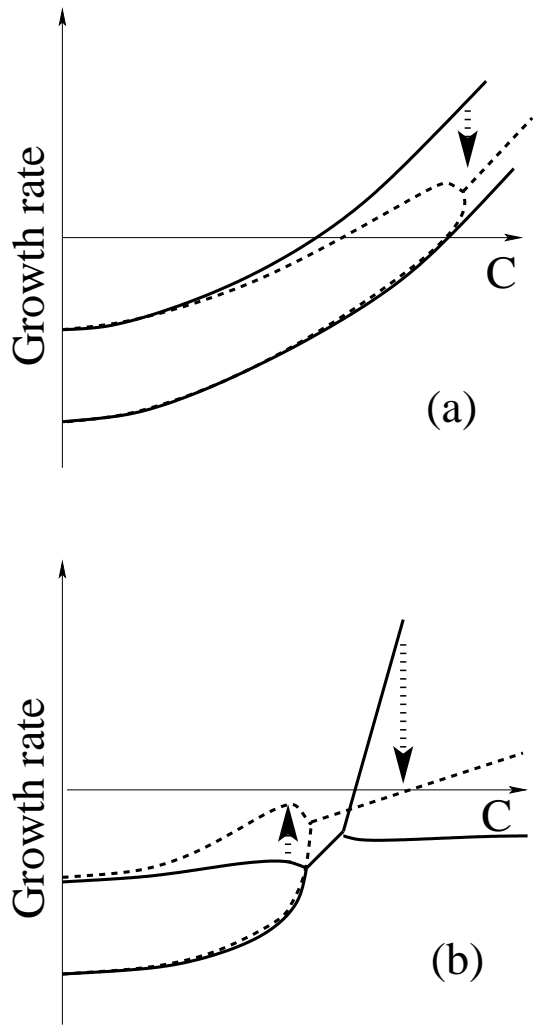}
\caption{The self-tuning mechanism of reversing dynamos. (a)
Case that there is no exceptional point or that it is above zero
in the kinematic regime.
(b) Case that the exceptional point is below zero in the kinematic regime.}
\end{figure}

At present, work is in progress to
support this rather intuitive  and numerically derived picture by
a paradigmatic analytical model.

In some respect, this self-tuning mechanism is similar to
the concept of {\it self-organized
criticality} \cite{SOC}, although it seems too early to
over-stress this resemblance.

\section{Time series and reversal duration}

The time evolution of the considered dynamos is, in the
noise-free case, determined by the position of the
local maximum of the growth rate curve
relative to the zero line. If the local maximum is
below zero, than we get a limit cycle in form
of an anharmonic oscillation,
the mechanism
of which has been described in detail above and in \cite{PRL}.
If the local maximum is situated well above the zero line,
then the system runs into
a stable fixed point. In case that the maximum is only
slightly above zero, than a hysteretic behaviour may occur,
in which the choice of the
stable fixed point or the limit cycle is controlled by
the strength of the pre-existing magnetic field.

A quantity which is of particular
geophysical relevance is the
duration of a reversal.
Here is not the place to discuss
the variety of different definitions, and
all questions concerning  the dependence of the apparent
reversal duration
on the site latitude \cite{CLEMENT}.
Since our model relies exclusively on the evolution of
the dipole field, we cannot
use any definitions based on directional changes.
For that reason we had employed in \cite{PRL} a
''working definition'' of a reversal duration,
based on the period during  which the modulus
of the magnetic field is smaller than a certain
percentage of the amplitude of the (anharmonic)
oscillation. For a value of 25 percent
(which was motivated by demanding the dipole field
to decay until this value in order that the non-dipole field
can become dominant)
we obtained a reversal duration of 0.15 $\tau_{diff}$.
For an assumed diffusion time of
200 kyr, this would amount to 30 kyr.
\begin{figure}
\includegraphics[width=7cm]{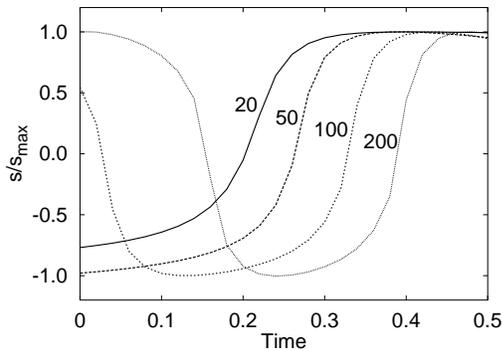}
\caption{Details of the reversal process for $\alpha_{kin}=6/3 \; C \; r^3$
with the particular values $C=$ 20, 50, 100, and 200.
The time scale is the
diffusion
time, which is approximately 200 kyr for the Earth. Evidently, the
reversal duration decreases significantly with increasing $C$ and
reaches values of $\sim$ 10 kyr for large $C$.}
\end{figure}

Now we would like to know how this reversal duration
changes with increasing magnetic Reynolds number, in particular for
highly supercritical dynamos.
In Fig. 8 we plot the details of the reversal process
for the third example of the previous section,
$\alpha_{kin}(r)=6/3 \; C\; r^3$.
For $C$ we have chosen the four values 20, 50, 100, and 200.
Whatever the
exact definition of a reversal might be, the tendency is
clear: with increasing $C$,
the reversal duration decreases significantly.
This has to do with the fact that during the reversal the
magnetic field gets weak and the
$\alpha$ profile comes close to its unquenched,
kinematic shape for which
extremely high instantaneous growth rates and frequencies
(only in the oscillatory branch)
occur.
These are responsible for a very fast reversal
process. If we would define the reversal duration
as the period during which the
field is between $\pm$ 50 percent of the
oscillation amplitude, we
would get for $C=200$ a time span of
0.03 $\tau_{diff}$, corresponding to
6 kyr. The corresponding time for $C=20$ is
0.09 $\tau_{diff}$, i.e. 18 kyr.
Evidently, we can get a realistic time scale
without taking resort to
turbulent resistivity, as it was done in other reversal models \cite{HOYNG}.

So far, we have considered the noise-free case.
As already remarked, the role of noise
depends on the position of the
local maximum of the
growth rate curve relative to the zero line.
If it is above zero, the dynamo is usually at the fixed point
characterized by a strong magnetic field, then
the noise can trigger
transitions to the unstable fixed point at the right of the
local maximum, from where a reversal process
can start. However, although being unstable, this
fixed point can hold the system for a while
(since the growth rate is
zero there)
making a second peak in the field strength histogram
possible \cite{PRL}.
If the local maximum is below zero, making the dynamo
undergo anharmonic oscillations with a weak field amplitude,
then the noise can make it
jump time by time to the strong field state.

Anyway, the noise will soften the differences
between the two regimes with the
local maximum  below or above zero. In either case,
there will be
noise-strength dependent exchange between the
fixed point state with a strong field and the limit cycle with a
weak field.

In Figs. 9 and 10 this is illustrated by some typical time
series for the four dynamo examples considered in the
previous section. In all
cases we have chosen
$C=50$. The noise intensity,
$D$ has been set to 4, apart from the first
example
with $\alpha_{kin}=C$ for which a value of
$D=7$ has been chosen to provoke any  reversal at all.
\begin{figure}
\includegraphics[width=7cm]{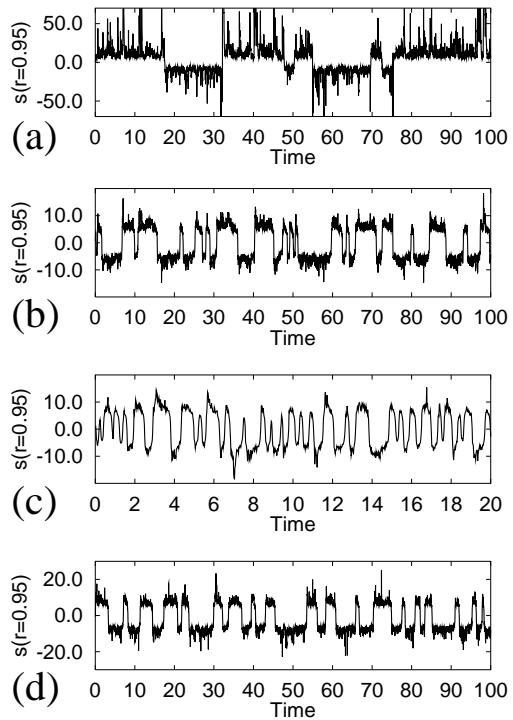}
\caption{Typical time series for the inclusion of noise.
(a) $\alpha_{kin}(r)= C$.
(b) $\alpha_{kin}(r)= 5/3 \; C \;r^2$
(c) $\alpha_{kin}(r)= 6/3 \;C \; r^3$
(d) $\alpha_{kin}(r)= 5.75/3 \; C \;(1-6 r^2+5 r^4)$
In all cases, we have chosen $C=50$. In the case (a), we have set $D=7$,
in all other cases (b-d) $D=4$.}
\end{figure}

\begin{figure}
\includegraphics[width=7cm]{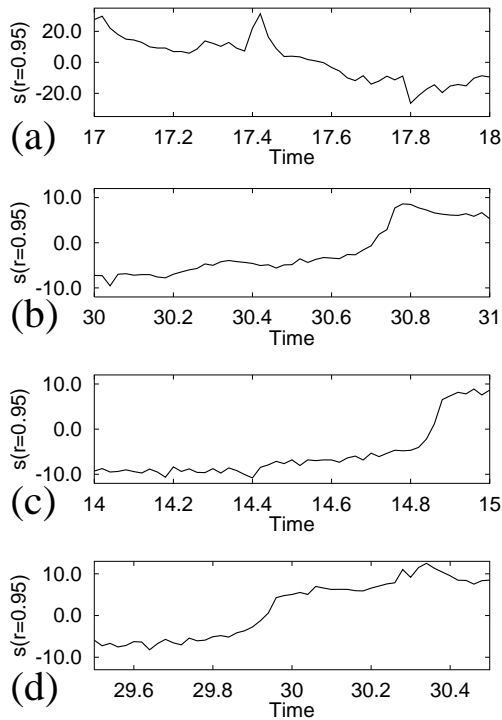}
\caption{Same as Fig. 9, but zoomed into time periods were
one reversal occurs.}
\end{figure}

Apart from details concerning the mean reversal
rate and the transient excursions to rather high values,
all the signals share the property that there are long time
intervals without reversals and
rather fast reversal processes.

A last remark concerns the
long-standing problem of positive or
negative correlation
of energy supply and interval length.
Our model is not able to solve this problem, but it
can illustrate its intricacy.
The first point is that any additional
energy supply will increase {\it both} the
magnetic Reynolds number $C$
{\it and} the noise level $D$.
Even if considered separately from $D$,
the influence of $C$
depends on whether it will lift or lower the local
maximum with respect to the zero line.
We have seen that both can happen, but for
high supercriticality there is a general tendency
of lowering the local maximum.
This would imply a negative
correlation of energy supply
and interval length.
If considering the influence of $D$
separately from
$C$, we get also a complicated behaviour with
a tendency toward a negative correlation
of $D$ and the interval length, but with
a positive correlation for the case that the local
maximum is below zero and the noise is weak (cf.
Fig. 7 of \cite{PRL}.
Taken all together we get a very complex picture
with  no definite answer.

\section{Conclusions}

In \cite{PRL} we had analyzed a reversal scenario that
relies heavily on the existence
of an exceptional point of the non-selfadjoint
dynamo operator.
We had shown that
reversals can a) be asymmetric, b) yield a positive
correlation of dynamo strength
and interval length, and c)  show a pronounced bimodal
field distribution. All three features have been
recently discussed
as being typical for reversals of the Earth's magnetic field.

The apparent  weakness of this model, the necessity of
{\it fine-tuning} the
spatial structure of the dynamo source in order to bring the
exceptional point
and its associated local growth rate maximum
close to the zero line,
has been
overcome in the present paper.
We have shown that
the back-reaction of the magnetic field has a strong
tendency to drive the dynamo into a state where the
indicated spectral
conditions for reversals are indeed fulfilled.

This shows that  dynamos with a high, but not
necessarily extreme, supercritical magnetic Reynolds
number are very prone to reversals. Hence, the proposed
reversal
scenario which might look contrived and hardly realistic
from a purely kinematic point of view,
becomes rather natural when seen from the side of the
saturation process.
The artificial fine-tuning for the
regime of slightly supercritical dynamos is
replaced by a
{\it self-tuning} saturation into reversal prone states
when it comes to highly supercritical dynamos.

The question remains if the Earth's dynamo is indeed
highly supercritical. A first estimate of the
magnetic Reynolds number,
based on a length scale of 2000 km, a velocity
scale of 5 mm/s, and
a magnetic diffusivity of $\lambda:=(\mu \sigma)^{-1}
\sim 2$ m$^2$/s
\cite{ROGLA},
provides $Rm\sim 500$. At least, and apart from
all uncertainties, this number
does not exclude a highly supercritical state.
Hence, whatever the concrete flow field in the Earth might be,
it is not a surprise that the resulting dynamo is prone
to reversals.

\section*{Acknowledgments}
This work was supported by
Deutsche Forschungsgemeinschaft
in frame of SFB 609 and Grant No.  GE 682/12-2.


\begin{thebibliography}{99}
\bibitem{PRL} F. Stefani, G. Gerbeth, Asymmetric polarity reversals,
bimodal field distribution, and coherence resonance in a
spherically symmetric mean-field dynamo model, Phys. Rev. Lett. 94 (2005) 184506,
physics/0411050. 
\bibitem{MERR} R.T. Merrill, M.W. McElhinny, P.L. McFadden,
The Magnetic Field of the Earth, Academic Press, San Diego, CA, 1996.
\bibitem{VALET} J.-P. Valet, L. Meynadier, Geomagnetic field
intensity and reversals during the past 4 million years,
Nature 366 (1993) 234-238.
\bibitem{MEYN} L. Meynadier, J.-P. Valet, F.C. Bassinot, N.J.
Shackleton, Y. Guyodo,
Asymmetrical saw-tooth pattern of the geomagnetic field intensity
from equatorial sediments in the pacific and indian
oceans, Earth Planet. Sci. Lett. 126 (1994) 109-127.
\bibitem{BOGU} S.W. Bogue, H.A. Paul,
Distinctive field behavior following geomagnetic reversals,
Geophys. Res. Lett. 20 (1993) 2399-2402.
\bibitem{COX} A. Cox, Length of geomagnetic polarity intervals,
J. Geophys. Res. 73 (1968) 3247-3259.
\bibitem{TARDUNO} J. A. Tarduno, R. D. Cottrell,
A. V. Smirnov,
High geomagntic intensity during the
mid-Cretaceous from Thellier analysis of
single plagioclase crystals,
Science 291 (2001) 1779.
\bibitem{FULLER} M. Fuller, R. Weeks,
Geomagnetism - Superplumes and Superchrons,
Nature 356 (1992) 16-17.
\bibitem{LARSON} R.L. Larson, The midcretacious superplume episode,
Scient. Amer. 272 (2) (1995) 82-86.
\bibitem{VOGT} P.R. Vogt, Changes in geomagnetic
reversal frequency at times of tectonic
change: evidence for coupling between core and
upper mantle
processes,
Earth Planet. Sci. Lett. 25 (1975) 313-321.
\bibitem{LARSOLS} R.L. Larson, P. Olson,
Mantle plumes control magnetic reversal frequency,
Earth Planet. Sci. Lett. 107 (1991) 437-447.
\bibitem{LOPER} D.E. Loper, On the correlation
between mantle
plume flux and the frequency of reversals of the
geomagnetic
field,
Geophys. Res. Lett. 19 (1992) 25-28.
\bibitem{LOMC} D.E. Loper, Mantle plumes and the
periodicity
of magnetic field reversals,
Geophys. Res. Lett. 13 (1986) 1525-1528.
\bibitem{OLSLEE} P. Olson, V. Lee Hagee,
Geomagnetic polarity reversals, transition field
structure, and convection in the outer core,
J. Geophys. Res. 95 (1990), 4609-4620.
\bibitem{PERRIN} M. Perrin, V.P. Shcherbakov,
Paleointensity of the Earth's magnetic field
for the past 400 Ma: Evidence for a dipole structure
during the mesozoic low, J. Geomagn. Geoelectr.
49 (1997) 601-614.
\bibitem{SHCH} V.P. Shcherbakov, G.M.
Solodovnikov, N.K. Sycheva,
Variations in the geomagnetic dipole during
the past 400 million years
(volcanic rocks), Izvestiya, Physics of the
Solid Earth 38 No. 2 (2002) 113-119.
\bibitem{HELLER} R. Heller, R. T. Merrill,
P. L. McFadden,
The two states of paleomagnetic field
intensities for the past 320 million years,
Phys. Earth Planet. Inter.
135 (2003)  211-223.
\bibitem{GLRO} G.A. Glatzmaier, P.H. Roberts,
A 3-dimensional convective dynamo solution with rotating
and finitely conducting inner core and mantle,
Phys. Earth Planet. Inter. 91 (1995) 63-75.
\bibitem{GLRO-ANDCO} G.A. Glatzmaier,
Geodynamo simulations - How realistic are they?
Annu. Rev. Earth Planet. Sci. 30 (2002) 237-257.
\bibitem{MININNI} P.D. Mininni, D. C. Montgomery, A.G. Pouquet,
A numerical study of the alpha model for two-dimensional
magnetohydrodynamic turbulent flows,
Phys. Rev. E. 71 (2005) 046304.
\bibitem{GIES} A. Giesecke, U. Ziegler, G. R\"udiger,
Geodynamo $\alpha$-effect derived from box simulations of
rotating magnetoconvection,
Phys. Earth Planet. Inter. 152 (2005) 90-102,
astro-ph/0410729.
\bibitem{SCHRINNER} M. Schrinner, K.-H. R\"adler, D. Schmitt, M.
Rheinhardt, U. Christensen,
Mean-field view on rotating magnetoconvection and a geodynamo model,
Astron. Nachr. 326 (2005) 245-249.
\bibitem{RIKI} T. Rikitake, Oscillations of a system of disk dynamos,
Proc. Cambridge Phil. Soc. 54 (1958) 89-105
\bibitem{FRANCK}
F. Plunian, P. Marty, A. Alemany,
Chaotic behaviour of the Rikitake dynamo with symmetric mechanical friction and azimuthal currents,
Proc. R. Soc. London, Ser. A 454 (1995) 1835-1842.
\bibitem{HOYNG} P. Hoyng, M. A. J. H. Ossendrijver, D. Schmidt,
The geodynamo as a bistable oscillator,
Geophys. Astroph.
Fluid Dyn. 94 (2001) 263-314.
\bibitem{SCHMITT}
D. Schmitt, M.A.J.H. Ossendrijver, P. Hoyng,
Magnetic field reversals and secular variation in a bistable geodynamo model,
Phys. Earth Planet. Inter. 125 (2001) 119-124.
\bibitem{HOYNG2} P. Hoyng, D. Schmitt, M.A.J.H. Ossendrijver,
A theoretical analysis of the observed variability of the geomagnetic dipole field,
Phys. Earth Planet. Inter. 130 (2002) 143-157.
\bibitem{LORITO} S. Lorito, D. Schmitt, G. Consolini, P. De Michelis,
Stochastic resonance in a bistable geodynamo model,
Astron. Nachr. 326 (2005) 227-230.
\bibitem{CONSOLINI} G. Consolini, P. De Michelis,
Stochastic resonance in geomagnetic polarity reversals,
Phys. Rev. Lett. 90 (2003) 058501.
\bibitem{YAMAODA} T. Yamazaki and H. Oda, Orbital influence on Earth's magnetic field: 100,000-year periodicity in inclination, Science 295 (2002) 2435-2438.
\bibitem{WINKL} A.P. Roberts, M. Winklhofer, W. T. Liang, et al.,
Testing the hypothesis of orbital (eccentricity) influence on Earth's magnetic field,
Earth Planet. Sci. Lett.  216 (2003) 187-192.
\bibitem{RMP} A. Gailitis, O. Lielausis, E. Platacis, G. Gerbeth, F. Stefani,
Colloquium: Laboraty experiments on hydromagnetic dynamos, Rev. Mod. Phys. 74
(2002) 973-990.
\bibitem{PARKER} E.N. Parker, Generation of
magnetic fields in astrophysical bodies. IV. Solar and
terrestrial dynamos, Astrophys. J. 164 (1971) 491-509
\bibitem{YOSHI} H. Yoshimura, Z. Wang, and F. Wu,
Linear astrophysical dynamos in rotating spheres: mode transition
between steady and oscillatory dynamos as a function of dynamo strenght
and anisotropic turbulent diffusivity, Astrophys. J. 283 (1984) 870-878.
\bibitem{SAJO}
G.R. Sarson and C.A. Jones, A convection driven geodynamo reversal model
Phys. Earth Planet. Inter. 111 (1999) 3-20.
\bibitem{DEIN} W. Deinzer, H.-U. v. Kusserow, M. Stix,
Steady and oscillatory $\alpha-\omega$ dynamos
Astron. Astrophys. 36 (1974) 69-78.
\bibitem{DUDL} M.L. Dudley, R.W. James,
Time-dependent kinematic dynamos with stationary
flows,
Proc. R. Soc. London. A. 425 (1989) 407-429.
\bibitem{KATO} T. Kato, Perturbation Theory of Linear Operators,
Springer, Berlin, 1966.
\bibitem{CZECH1} U. G\"unther, F. Stefani, G. Gerbeth,
The MHD $\alpha^2$-dynamo, $\ZZ_2-$ graded
pseudo-Hermiticity, level
crossings and exceptional points of branching type,
Czech. J. Phys. 54
(2004) 1075-1089,
math-ph/0407015.
\bibitem{SEYR} A.P. Seyranian, O.N. Kirillov, A.A. Mailybaev,
Coupling of eigenvalues of complex matrices at diabolic and
exceptional points,
J. Phys. A 38 (2005) 1723-1740,
math-ph/0411024.
\bibitem{CZECH2} U. G\"unther, F. Stefani,
Third order spectral branch points in Krein space related setups:
PT-symmetric matrix toy model, MHD $\alpha^2$-dynamo, and
extended Squire equation, Czech. J. Phys. 55 (2005) in press,
math-ph/0506021.
\bibitem{BERRY} M.V. Berry, M. Wilkinson,
Diabolical points in the spectra of triangles,
Proc. R. Soc. Lond. A392 (1984) 15-43.
\bibitem{BRANKRAU}  A. Brandenburg,
F. Krause, R. Meinel, D. Moss, I. Tuominen,
The stability of nonlinear dynamos and the limited
role of kinematic grwoth rates,
Astron. Astrophys. 213 (1989) 411-422.
\bibitem{MEINBRAN} R. Meinel, A. Brandenburg,
Behaviour of highly supercritical $\alpha$-effect dynamos,
Astron. Astrophys. 238 (1990) 369-376.
\bibitem{PLASMA} A. Gailitis, O. Lielausis, E. Platacis,
G. Gerbeth, F. Stefani, Riga dynamo experiment and its theoretical
background, Phys. Plasmas  11 (2004) 2838-2843.
\bibitem{MOMOMO} A. Gailitis, O. Lielausis, G. Gerbeth, F. Stefani,
Dynamo experiments, in S. Molokov, R. Moreau, H.K. Moffatt (Eds.): ''Magnetohydrodynamics: evolution of ideas and trends'',
Berlin: Springer/Kluwer, 2005, to appear.
\bibitem{KRRA} F. Krause and K.-H. R\"adler, Mean-field Magnetohydrodynamics and
Dynamo Theory, Akademie-Verlag, Berlin, 1980.
\bibitem{NAKAKONO} T. Nakajima, M. Kono,
Kinematic dynamos associated with large scale fluid motions,
Geophys. Astrophys. Fluid Dyn. 60 (1991) 177-209.
\bibitem{OSZI} F. Stefani and G. Gerbeth,
Oscillatory mean-field dynamos with a spherically symmetric, isotropic helical turbulence parameter $\alpha$,
Phys. Rev. E 67 (2003) 027302.
\bibitem{OLSON} P. Olson, Geomagnetic polarity reversals in a turbulent
core, Phys. Earth Planet. Inter. 33 (1983) 260-274.
\bibitem{VANDERPOL} B. van der Pol, On relaxation oscillations,
Philos. Mag. 2 (1926) 978-992
\bibitem{SOC} P. Bak, C. Tang, K. Wiesenfeld,
Self-organized criticality, Phys. Rev. A 38 (1988) 364-374.
\bibitem{CLEMENT} B.M. Clement, Dependence of the
duration of geomagnetic polarity reversals on site latitude,
Nature 428 (2004) 637-640.
\bibitem{ROGLA} P.H. Roberts, G.A. Glatzmaier, Geodynamo theory and simulations,
Rev. Mod. Phys. 72 (2000) 1081-1123.
\end{thebibliography}
\end{document}